\documentclass[11pt]{article}
\usepackage{amsfonts}
\usepackage{a4,tikz}
\usepackage{geometry}
\usepackage{color} 
\usepackage{subfigure}
\makeatletter
\newcommand\ackname{Acknowledgements}
\if@titlepage
  \newenvironment{acknowledgements}{%
      \titlepage
      \null\vfil
      \@beginparpenalty\@lowpenalty
      \begin{center}%
        \bfseries \ackname
        \@endparpenalty\@M1
      \end{center}}%
     {\par\vfil\null\endtitlepage}
\else
  
\fi
\makeatother
\usepackage{amsmath}
\usepackage{amssymb}
\usepackage{amsthm}
\usepackage{mathrsfs}
\usepackage{tabularx}
\usepackage{eucal}
 \usepackage[usenames,dvipsnames]{pstricks}
 \usepackage{epsfig}
 \usepackage{pst-grad} 
 \usepackage{pst-plot} 
\renewcommand{\theequation}{\arabic{equation}}
\usepackage{dsfont}
\usepackage{amssymb,amsmath}
\usepackage{color}
\usepackage{accents}
\usepackage{texdraw}
\usepackage{eucal}
\usepackage{multirow}
\usepackage[utf8]{inputenc}
\usepackage{cancel}
\usetikzlibrary{ decorations.markings}
\usepackage{epic,epsfig}
\usepackage{graphicx}

\theoremstyle{definition}

\numberwithin{equation}{section}
\DeclareMathAccent{\wtilde}{\mathord}{largesymbols}{"65}
\DeclareMathAccent{\what}{\mathord}{largesymbols}{"62}

\def\m@th{\mathsurround=0pt}
\mathchardef\bracell="0365
\def\upbrall{$\m@th\bracell$}
\def\undertilde#1{\mathop{\vtop{\ialign{##\crcr
    $\hfil\displaystyle{#1}\hfil$\crcr
     \noalign
     {\kern1.5pt\nointerlineskip}
     \upbrall\crcr\noalign{\kern1pt
   }}}}\limits}
\def\m@th{\mathsurround=0pt}
\mathchardef\bracell="0365
\def\upbrall{$\m@th\bracell$}
\def\underhat#1{\mathop{\vtop{\ialign{##\crcr
    $\hfil\displaystyle{#1}\hfil$\crcr
     \noalign
     {\kern1.5pt\nointerlineskip}
     \upbrall\crcr\noalign{\kern1pt
   }}}}\limits}
\usepackage{pgf}
\usepackage{tikz}
\usepackage{cancel,xcolor}
\usepackage{verbatim}
\usepackage[utf8]{inputenc}
\usetikzlibrary{arrows,automata}
\usetikzlibrary{positioning}
\usepackage[toc,page]{appendix}

\def\theequation{\arabic{section}.\arabic{equation}}

\newcommand{\bblu}{\begin{color}{blue}}
\newcommand{\bred}{\begin{color}{red}}
\newcommand{\ecl}{\end{color}}

\newcommand{\be}{\begin{equation}}
\newcommand{\ee}{\end{equation}}
\newcommand{\bea}{\begin{eqnarray}}
\newcommand{\eea}{\end{eqnarray}}
\newcommand{\bse}{\begin{subequations}}
\newcommand{\ese}{\end{subequations}}

\title{Maxwell's relations as Hamilton's equations: a symplectic and variational framework}
\author{Sikarin Yoo-Kong \\
            \small {The Institute for Fundamental Study (IF),} \small\emph{Naresuan University, Phitsanulok, Thailand, 65000.}\\
		\small{sikariny@nu.ac.th} \\
	}
\begin{document}
\def\theequation{\arabic{section}.\arabic{equation}}

\newtheorem{thm}{Theorem}[section]
\newtheorem{lem}{Lemma}[section]
\newtheorem{defn}{Definition}[section]
\newtheorem{ex}{Example}[section]
\newtheorem{rem}{}
\newtheorem{criteria}{Criteria}[section]
\newcommand{\ra}{\rangle}
\newcommand{\la}{\langle}
\maketitle

\begin{abstract}
We develop a geometric framework in which classical thermodynamics is
reformulated as a Hamiltonian dynamical system.  An explicit canonical
mapping $(q,p,t,H)\leftrightarrow(V,-P,S,-T)$ identifies Maxwell's
relations as the characteristic equations of the thermodynamic
Poincar\'e--Cartan one-form, in direct parallel with Hamilton's
equations of motion.  Treating thermodynamic potentials as action
functionals, a variational principle recovers both the Maxwell
relations and the thermodynamic constraints (adiabaticity,
isothermality) as conserved first integrals.  For adiabatic processes
in an ideal gas, this yields explicit second-order ordinary
differential equations for $V(T)$ and $P(T)$, each governed by a
temperature-dependent Lagrangian.  Assembling the component Lagrangians
into a multi-parameter Lagrangian 1-form
$\mathcal{L}$, we prove
that the closure condition $d\mathcal{L}=0$ holds on the solution
manifold, establishing that path-independence of thermodynamic state
functions is a geometric consequence of multi-time integrability rather
than an independent postulate.
\end{abstract}

\section{Introduction}
\label{sec:intro}
 
The mathematical foundation of classical thermodynamics has undergone a
profound transformation since the seminal work of Carath\'eodory
\cite{Caratheodory1909} and the subsequent development of geometric
thermodynamics by Hermann \cite{Hermann1973} and Mrugała
\cite{Mrugala1978,Mrugala1991}.  The traditional view of thermodynamics
as a collection of empirical laws has evolved into a formulation rooted
in differential geometry, where equilibrium states are points on a
manifold and quasi-static processes are curves connecting them
\cite{Callen1985}.  Despite this elegance, a conceptual gap remains:
the derivation of continuous thermodynamic trajectories from a unified
variational principle.  While the conjugate pairs $(P,V)$ and $(T,S)$
strikingly mirror the canonical pairs $(q,p)$ of classical mechanics
\cite{Arnold1989}, thermodynamic evolution is traditionally described
only as a sequence of quasi-static equilibrium processes, with no
underlying action principle generating the path.  This raises a natural
question: can the evolution of a thermodynamic system between
equilibrium states be described as a continuous flow generated by an
action-like potential on a symplectic manifold?
 \\
 \\
The geometric thermodynamics literature has approached related questions
through contact geometry \cite{Mrugala1978,Mrugala1991,Bravetti2015},
which is the natural setting for non-equilibrium extensions.  The
present paper works instead within symplectic and presymplectic
geometry, making the Hamiltonian analogy quantitative through explicit
Lagrangians, equations of motion, and a closed Lagrangian 1-form.  The
Lagrangian 1-form framework was introduced in \cite{YooKong2011}, and
its application to equilibrium thermodynamics, specifically the
identification of the closure condition $d\mathcal{L}=0$ as the
integrability condition for thermodynamic state functions, was
established in \cite{YooKong2025}.  The present paper extends that
foundation by providing the full geometric underpinning: an explicit
canonical correspondence between Hamiltonian mechanics and
thermodynamics, a variational derivation of Maxwell's relations as
Euler--Lagrange equations, and explicit temperature-dependent
Lagrangians for adiabatic processes.
\\
\\
The main results are as follows.  We construct a canonical mapping
$(q,p,t,H) \leftrightarrow (V,-P,S,-T)$ under which Maxwell's relations
emerge as the characteristic equations of the thermodynamic
Poincar\'e--Cartan one-form, and we confirm the correspondence through
matching Poisson brackets $\{P,V\}=1$, $\{S,T\}=-1$.  Treating $U$
and $G$ as action functionals, we show that their stationarity
conditions reproduce the Maxwell relations, with the adiabatic and
isothermal constraints arising automatically as conserved first
integrals.  For an ideal gas, this yields uncoupled second-order ODEs
for $V(T)$ and $P(T)$ with explicit Lagrangians $L_V$ and $L_P$.
Finally, the Lagrangian 1-form $\mathcal{L} = \mathcal{L}_U\,d\xi_1 +
\mathcal{L}_G\,d\xi_2$ is shown to satisfy $d\mathcal{L}=0$ on the
solution manifold, establishing path-independence of state functions as
a consequence of multi-time integrability rather than an independent
postulate.
 \\
 \\
The paper is organised as follows.
Section~\ref{sec:hamiltonian} reviews Hamiltonian mechanics.
Section~\ref{sec:thermo} reviews equilibrium thermodynamics.
Section~\ref{sec:comparison} establishes the canonical correspondence.
Section~\ref{sec:evolution} derives thermodynamic equations of motion
and Lagrangians.
Section~\ref{sec:variational} develops the variational principles.
Section~\ref{sec:closure} proves the closure relation.
Section~\ref{sec:conclusion} summarises and outlines future directions.
\section{Hamiltonian mechanics: variational and geometric formulation}
\label{sec:hamiltonian}
 
In this section we briefly review the variational and differential-geometric
formulation of Hamiltonian mechanics that will serve as the mathematical
template for the thermodynamic construction developed below.  We emphasise
the distinction between the Hamiltonian action evaluated along a trajectory,
the Poincar\'e--Cartan one-form on extended phase space, the associated
two-form, and the local exact representation provided by Hamilton--Jacobi
theory.
 
\subsection{Hamiltonian variational principle}
 
Consider a classical mechanical system with $N$ degrees of freedom with
generalised coordinates $q = (q_1,\ldots,q_N)$, canonical momenta
$p = (p_1,\ldots,p_N)$, and Hamiltonian $H = H(q,p,t)$.  The action
functional is
\begin{equation}
  \label{eq:action_H}
  A[p,q] = \int_{t'}^{t''} dt \left[\sum_{i=1}^N p_i\,\frac{dq_i}{dt}
            - H(q,p,t)\right].
\end{equation}
Requiring the action to be stationary, $\delta A = 0$, under independent
variations $q\to q+\delta q$ and $p\to p+\delta p$ with fixed endpoints,
$\delta q_i(t')=\delta q_i(t'')=0$, yields Hamilton's equations of motion,
\begin{equation}
  \label{eq:Hamilton_eom}
  \frac{dq_i}{dt} = \frac{\partial H}{\partial p_i}, \qquad
  \frac{dp_i}{dt} = -\frac{\partial H}{\partial q_i}, \qquad
  i = 1,\ldots,N.
\end{equation}
Hamiltonian mechanics can therefore be regarded as a variational theory in
which the canonical pair $(q_i,p_i)$ evolves with respect to the parameter
$t$.
 
\subsection{Poincar\'e--Cartan one-form}
 
The Hamiltonian action admits a natural differential-geometric representation
on the extended phase space with coordinates $(q,p,t)$.  We introduce the
\emph{Poincar\'e--Cartan one-form}
\begin{equation}
  \label{eq:PC_form}
  \Theta_H = \sum_{i=1}^N p_i\,dq_i - H\,dt.
\end{equation}
Given a trajectory $\gamma : t \mapsto (q(t),p(t))$, the pullback to $\gamma$
satisfies $dq_i = (dq_i/dt)\,dt$, so
\begin{equation}
  \label{eq:action_pullback}
  A[\gamma] = \int_\gamma \Theta_H
            = \int_{t'}^{t''} \left(\sum_{i=1}^N p_i\,\frac{dq_i}{dt}
              - H\right) dt.
\end{equation}
The exterior derivative of $\Theta_H$ defines the presymplectic two-form,
\begin{equation}
  \label{eq:presymplectic}
  \Omega_H = d\Theta_H = \sum_{i=1}^N dp_i \wedge dq_i - dH \wedge dt.
\end{equation}
Because $d^2 = 0$ it follows that $d\Omega_H = 0$, i.e.\ $\Omega_H$ is
closed.  Importantly, however, $\Omega_H$ is not generally exact: $\Theta_H$
is a geometrically defined one-form on extended phase space and is not, in
general, the differential of a globally defined scalar function.  The
distinction between closedness and exactness of $\Theta_H$ will be central
when we construct the analogous thermodynamic differential forms below.
 \\
 \\
Hamiltonian trajectories admit a concise geometric characterisation.  Let
\begin{equation}
  \label{eq:tangent_vec}
  \mathbf{X} = \frac{\partial}{\partial t}
      + \sum_{i=1}^N \left(\frac{dq_i}{dt}\,\frac{\partial}{\partial q_i}
        + \frac{dp_i}{dt}\,\frac{\partial}{\partial p_i}\right)
\end{equation}
be the tangent vector to a parametrised trajectory.  Physical trajectories
are the characteristic curves of $\Omega_H$, i.e.\ the integral curves of
the vector field $X$ satisfying
\begin{equation}
  \label{eq:char_curves}
  \iota_\mathbf{X} \Omega_H = 0,
\end{equation}
where $\iota_\mathbf{X}$ denotes interior multiplication by $\mathbf{X}$.  Substituting
\eqref{eq:presymplectic} and \eqref{eq:tangent_vec} into \eqref{eq:char_curves}
recovers Hamilton's equations \eqref{eq:Hamilton_eom}, providing the geometric
interpretation of Hamiltonian dynamics as the characteristic flow of the
Poincar\'e--Cartan structure.
 
\subsection{Hamilton--Jacobi formulation and exact differentials}
 
A different but related structure arises when the Poincar\'e--Cartan
one-form admits a \emph{local} exact representation.  Suppose that on a
suitable local neighbourhood of the solution manifold there exists a
generating function $A[q,t]$ such that
\begin{equation}
  \label{eq:HJ_exact}
  dA = \Theta_H = \sum_{i=1}^N p_i\,dq_i - H\,dt.
\end{equation}
Matching coefficients gives
\begin{equation}
  \label{eq:HJ_momenta}
  p_i = \frac{\partial A}{\partial q_i}, \qquad
  H   = -\frac{\partial A}{\partial t}.
\end{equation}
The nilpotency $d^2 A = 0$ is then satisfied identically, and in local
coordinates the equality of mixed partial derivatives of $A$ yields the
\emph{Hamilton--Jacobi equation},
\begin{equation}
  \label{eq:HJ_eq}
  \frac{\partial A}{\partial t}
  + H\!\left(q,\frac{\partial A}{\partial q},t\right) = 0.
\end{equation}
It is important to distinguish \eqref{eq:HJ_exact} from the definition of
$\Theta_H$ in \eqref{eq:PC_form}.  The Poincar\'e--Cartan one-form is a
geometrically defined object on extended phase space. Equation \eqref{eq:HJ_exact}
asserts the additional, and generally only local, condition that $\Theta_H$
equals the exact differential of a scalar.  Consequently $d\Theta_H =
\Omega_H$ is generally non-zero even when the local exact representation
\eqref{eq:HJ_exact} exists.
 
\subsection{Extended canonical structure}
 
The role of time can be incorporated into the canonical formalism
symmetrically with the other coordinates by introducing the
\emph{extended phase space} with coordinates $(q_i,p_i,t,p_t)$, where $p_t$
is the momentum conjugate to $t$.  The extended symplectic form is
\begin{equation}
  \label{eq:extended_symp}
  \Omega_{\mathrm{Ext}}
  = \sum_{i=1}^N dp_i \wedge dq_i + dp_t \wedge dt,
\end{equation}
with canonical Poisson brackets $\{q_i,p_j\} = \delta_{ij}$,
$\{t,p_t\} = 1$, all other elementary brackets vanishing.  Ordinary
Hamiltonian dynamics is recovered by restricting to the constraint surface
$p_t + H(q,p,t) = 0$; on this surface $p_t = -H$, so $t$ and $H$ enter the
formalism as a canonically conjugate pair.  This extended canonical structure
should be distinguished from the ordinary phase-space Poisson structure, in
which $t$ is an external evolution parameter rather than a canonical
coordinate.
 \\
 \\
The present construction anticipates the thermodynamic framework developed in
sections~\ref{sec:comparison}--\ref{sec:closure}: the two thermodynamic
parameters $(\xi_1,\xi_2)$ introduced there play exactly the role of
extended-time-like coordinates, and the conjugate pairings $(T,S)$ and
$(-P,V)$ mirror the extended canonical structure of \eqref{eq:extended_symp}.
%
%
%
%
%
%
%
%
%
%
%
%
%
%
\section{A short review on thermodynamics}
\label{sec:thermo}
 
In this section we review the mathematical framework of equilibrium
thermodynamics, focusing on state variables, differential forms,
thermodynamic potentials, and the emergence of Maxwell relations.
The presentation is organised to make the structural parallel with
section~\ref{sec:hamiltonian} explicit: the fundamental thermodynamic
relation plays the role of the Poincar\'e--Cartan one-form, exactness
of the potentials replaces the Hamilton--Jacobi generating-function
condition, and the Maxwell relations arise from the same nilpotency
identity $d^2 = 0$ that produces Hamilton's equations in the mechanical
setting.
 
\subsection{State variables and the fundamental relation}
 
The equilibrium state of a simple compressible system is completely
specified by two independent macroscopic variables.  We work with
entropy $S$ and volume $V$ as the primary extensive variables, with
the conjugate intensive quantities temperature $T$ and pressure $P$
defined through partial derivatives of the internal energy $U(S,V)$.
 \\
 \\
For a quasi-static process the first and second laws combine to give
the \emph{fundamental thermodynamic relation},
\begin{equation}
  \label{eq:first_law}
  dU = T\,dS - P\,dV,
\end{equation}
from which the intensive variables are read off as
\begin{equation}
  \label{eq:intensive}
  T = \left.\frac{\partial U}{\partial S}\right|_V, \qquad
  P = -\left.\frac{\partial U}{\partial V}\right|_S.
\end{equation}
The pairs $(T,S)$ and $(-P,V)$ are the thermodynamic analogues of the
canonical pairs $(p_i,q_i)$ in Hamiltonian mechanics: each consists of
an intensive (momentum-like) and an extensive (coordinate-like) variable.
This identification will be made precise in section~\ref{sec:comparison}.
 \\
 \\
Because $U$ is a state function, single-valued and path-independent, its
differential $dU$ is exact.  For any quasi-static cycle $\Gamma$ connecting
equilibrium states $A$ and $B$,
\begin{equation}
  \label{eq:path_indep}
  \int_\Gamma dU = U(B) - U(A),
\end{equation}
independently of the path chosen.  This is the thermodynamic counterpart
of the Hamilton--Jacobi condition \eqref{eq:HJ_exact}: just as $dA =
\Theta_H$ requires $\Theta_H$ to be locally exact, the exactness of $dU$
encodes the single-valued character of the internal energy.
 
\subsection{Thermodynamic potentials and Legendre transformations}
 
To describe processes under varying natural constraints it is convenient
to pass to alternative potentials obtained by Legendre transformation of
$U(S,V)$.  Each transformation exchanges one conjugate pair for the other,
yielding a new potential whose natural variables are adapted to the
relevant experimental conditions.
 
\medskip
\noindent\textbf{Enthalpy} $H(S,P)$:
\begin{equation}
  \label{eq:enthalpy}
  H = U + PV \implies dH = T\,dS + V\,dP.
\end{equation}
 
\medskip
\noindent\textbf{Helmholtz free energy} $F(T,V)$:
\begin{equation}
  \label{eq:helmholtz}
  F = U - TS \implies dF = -S\,dT - P\,dV.
\end{equation}
 
\medskip
\noindent\textbf{Gibbs free energy} $G(T,P)$:
\begin{equation}
  \label{eq:gibbs}
  G = H - TS \implies dG = -S\,dT + V\,dP.
\end{equation}
 
\medskip\noindent
All four potentials $U$, $H$, $F$, $G$ are exact differentials, so their
line integrals between any two equilibrium states are path-independent,
in exact parallel with \eqref{eq:path_indep}.
 
\subsection{Maxwell relations from exactness}
 
Because each thermodynamic potential is a state function, its second
mixed partial derivatives commute.  This is the thermodynamic
manifestation of the nilpotency identity $d^2 = 0$ used in
section~\ref{sec:hamiltonian}: applying $d$ twice to any exact scalar
gives zero.  Applied to each potential, this yields four \emph{Maxwell
relations}.
 
\medskip
\noindent\textbf{From} $dU = T\,dS - P\,dV$:
\begin{equation}
  \label{eq:maxwell_U}
  \left.\frac{\partial T}{\partial V}\right|_S
  = -\left.\frac{\partial P}{\partial S}\right|_V.
\end{equation}
 
\medskip
\noindent\textbf{From} $dH = T\,dS + V\,dP$:
\begin{equation}
  \label{eq:maxwell_H}
  \left.\frac{\partial T}{\partial P}\right|_S
  = \left.\frac{\partial V}{\partial S}\right|_P.
\end{equation}
 
\medskip
\noindent\textbf{From} $dF = -S\,dT - P\,dV$:
\begin{equation}
  \label{eq:maxwell_F}
  \left.\frac{\partial S}{\partial V}\right|_T
  = \left.\frac{\partial P}{\partial T}\right|_V.
\end{equation}
 
\medskip
\noindent\textbf{From} $dG = -S\,dT + V\,dP$:
\begin{equation}
  \label{eq:maxwell_G}
  -\left.\frac{\partial S}{\partial P}\right|_T
  = \left.\frac{\partial V}{\partial T}\right|_P.
\end{equation}
 
\noindent
Each relation follows from the same argument: if $d\Phi = X\,da + Y\,db$
for a state function $\Phi(a,b)$, then $d^2\Phi = 0$ forces
$\partial X/\partial b = \partial Y/\partial a$.  This is structurally
identical to the consistency condition that produced the Hamilton--Jacobi
equation \eqref{eq:HJ_eq} from $d^2 A = 0$.
 
\subsection{Symplectic structure of thermodynamic differential forms}
 
The parallel with Hamiltonian mechanics extends to the differential-geometric
level.  Define the \emph{thermodynamic one-form}
\begin{equation}
  \label{eq:thermo_1form}
  \omega_U = -P\,dV + T\,dS,
\end{equation}
which is the direct analogue of the Poincar\'e--Cartan one-form
$\Theta_H = \sum p_i\,dq_i - H\,dt$ of section~\ref{sec:hamiltonian},
with $(T,S) \leftrightarrow (p_i,q_i)$ and $(-P,V) \leftrightarrow
(p_j,q_j)$.  Taking the exterior derivative yields the
\emph{thermodynamic two-form},
\begin{equation}
  \label{eq:thermo_2form}
  \Omega_U = d\omega_U = dV \wedge dP - dS \wedge dT,
\end{equation}
which is closed by nilpotency, $d\Omega_U = d^2\omega_U = 0$, in exact
parallel with $d\Omega_H = 0$ for the presymplectic two-form
\eqref{eq:presymplectic}.
 \\
 \\
The physical content of $\Omega_U$ is made vivid by integrating over a
closed quasi-static cycle.  Since $U$ is a state function,
\begin{equation}
  \label{eq:cycle_integral}
  \oint dU = \oint \omega_U = \oint (T\,dS - P\,dV) = 0,
\end{equation}
and by Stokes' theorem
\begin{equation}
  \label{eq:stokes}
  \iint dT \wedge dS = \iint dP \wedge dV.
\end{equation}
This identity expresses the well-known result that the area enclosed by a
quasi-static cycle in the $(T,S)$ plane equals the net mechanical work in the
$(P,V)$ plane (A direct consequence of the symplectic structure
\eqref{eq:thermo_2form}).  The thermodynamic phase space therefore carries an
intrinsic symplectic geometry that mirrors the phase-space structure of
Hamiltonian mechanics, a correspondence that will be made explicit through a
canonical Poisson bracket comparison in section~\ref{sec:comparison}.
\section{Canonical correspondence between Hamiltonian mechanics
         and thermodynamics}
\label{sec:comparison}
 
\begin{figure}[h]
\centering
\includegraphics[width=0.75\linewidth]{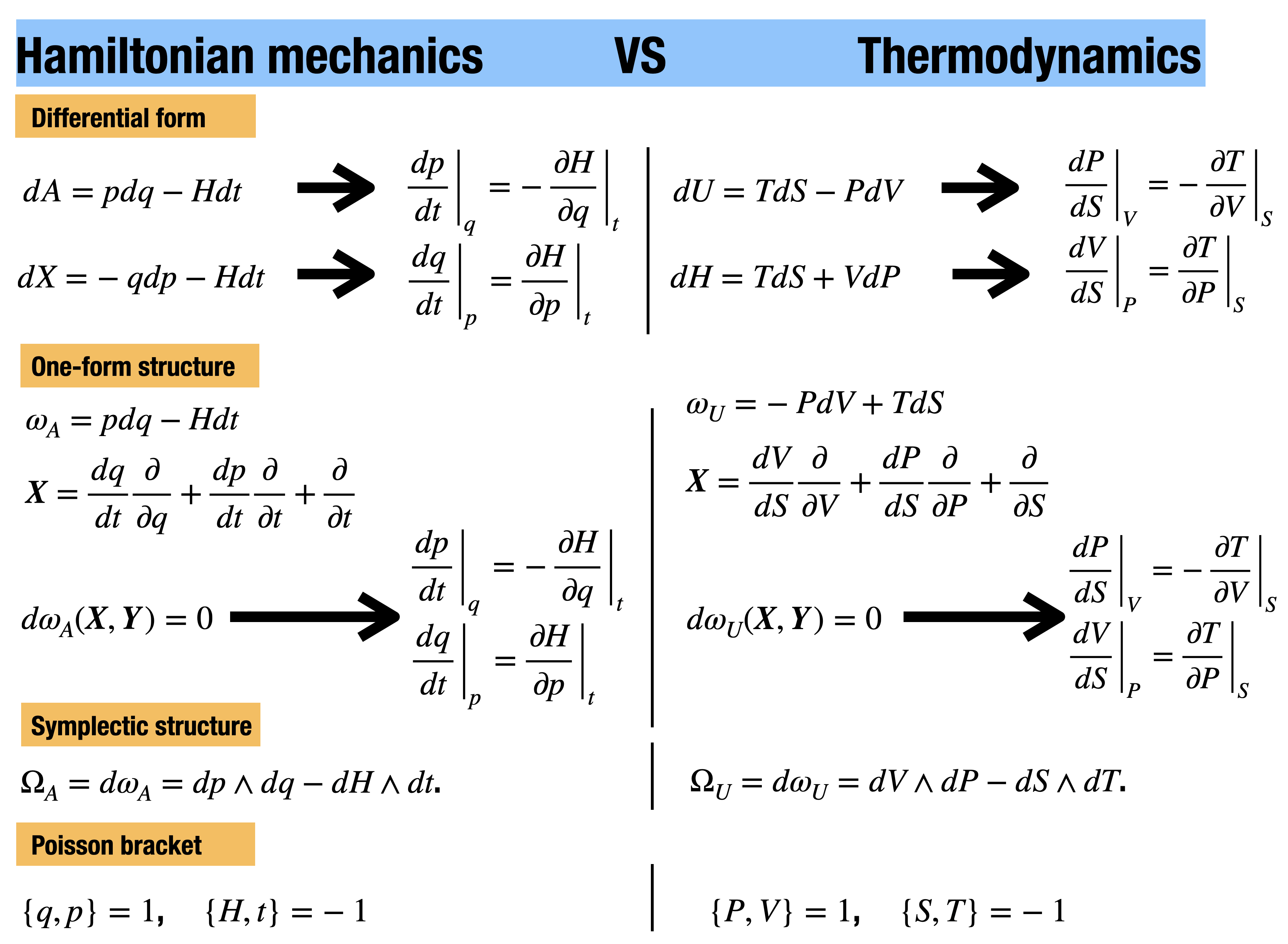}
\caption{\label{HVT} The comparison structure between Hamiltonian mechanics and thermodynamics.}
\end{figure}

Having established the geometric structure of Hamiltonian mechanics in
section~\ref{sec:hamiltonian} and of equilibrium thermodynamics in
section~\ref{sec:thermo}, we now make the structural analogy between the
two theories precise.  The comparison is summarised in
figure~\ref{HVT}-the four subsections below derive each
row explicitly.
\subsection{Canonical variable correspondence}
 
The fundamental differential relations of both theories have identical algebraic form.  In mechanics, the action differential and its
variation yield
\begin{equation}
  \label{eq:mech_diff}
  dA = \sum_{i=1}^N p_i\,dq_i - H\,dt,
  \qquad
  \left.\frac{dq_i}{dt}\right|_p = \frac{\partial H}{\partial p_i}.
\end{equation}
\begin{equation}
  \label{eq:mech_diff}
  dX = -\sum_{i=1}^N q_i\,dp_i - H\,dt,
  \qquad
  \left.\frac{dp_i}{dt}\right|_q = -\frac{\partial H}{\partial q_i}.
\end{equation}
In thermodynamics the internal-energy differential and its partial
derivatives yield
\begin{equation}
  \label{eq:thermo_diff}
  dU = T\,dS - P\,dV,
  \qquad
  \left.\frac{\partial P}{\partial S}\right|_V = -\frac{\partial T}{\partial V}.
\end{equation}
\begin{equation}
  \label{eq:thermo_diff}
  dH = T\,dS +V\,dP,
  \qquad
  \left.\frac{\partial V}{\partial S}\right|_P = \frac{\partial T}{\partial P}.
\end{equation}
Comparing \eqref{eq:mech_diff} and \eqref{eq:thermo_diff} identifies the
canonical correspondence
\begin{equation}
  \label{eq:canonical_map}
  \begin{aligned}
    q_i   &\longleftrightarrow V,  \qquad &
    p_i   &\longleftrightarrow -P, \\
    t     &\longleftrightarrow S,  \qquad &
    H     &\longleftrightarrow -T.
  \end{aligned}
\end{equation}
These identifications are not merely mnemonic: they assert that $(V,-P)$
and $(S,T)$ are thermodynamic \emph{canonical pairs} in the same sense
that $(q_i,p_i)$ are mechanical canonical pairs.
 
\subsection{One-form structure and characteristic equations}
 
Each theory is organised around a fundamental one-form whose
characteristic curves encode the equations of motion (mechanics) or
equilibrium relations (thermodynamics).
 \\
 \\
In mechanics, the Poincar\'e--Cartan one-form
$\Theta_H = \sum p_i\,dq_i - H\,dt$ generates
Hamilton's equations as the characteristic curves of $\Omega_H =
d\Theta_H$ via $\iota_\mathbf{X} \Omega_H = 0$.
 \\
 \\
In thermodynamics, the energy one-form
$\omega_U = -P\,dV + T\,dS$ plays the
identical role.  Define the thermodynamic tangent vector
\begin{equation}
  \label{eq:thermo_tangent}
  \mathbf{X} = \frac{dV}{dS}\,\frac{\partial}{\partial V}
             + \frac{dP}{dS}\,\frac{\partial}{\partial P}
             + \frac{\partial}{\partial S}
\end{equation}
along a quasi-static path parametrised by $S$.  Requiring $\iota_{\mathbf{X}}
\Omega_U = 0$, where $\Omega_U = d\omega_U$ is the thermodynamic two-form
\eqref{eq:thermo_2form}, reproduces the equilibrium conditions
\begin{equation}
  \label{eq:thermo_char}
  \left.\frac{\partial P}{\partial S}\right|_V = -\frac{\partial T}{\partial V},
  \qquad
  \left.\frac{\partial V}{\partial S}\right|_P = \frac{\partial T}{\partial P},
\end{equation}
which are precisely the Maxwell relations \eqref{eq:maxwell_U} and
\eqref{eq:maxwell_H}.  The Maxwell relations therefore arise as the
characteristic equations of the thermodynamic Poincar\'e--Cartan
structure, in exact parallel with Hamilton's equations in mechanics.
 
\subsection{Symplectic two-form correspondence}
 
Taking the exterior derivative of each fundamental one-form yields a
closed two-form that encodes the geometry of the respective phase space:
\begin{align}
  \label{eq:symp_mech}
  \Omega_H &= d\Theta_H = \sum_{i=1}^N dp_i \wedge dq_i - dH \wedge dt
           & &\text{(mechanical)},\\[4pt]
  \label{eq:symp_thermo}
  \Omega_U &= d\omega_U = dV \wedge dP - dS \wedge dT
           & &\text{(thermodynamic)}.
\end{align}
Both two-forms are closed by nilpotency ($d^2 = 0$) and non-degenerate on
their respective even-dimensional subspaces, establishing a symplectic
structure in each case.  Under the canonical map \eqref{eq:canonical_map},
the mechanical symplectic element $dp \wedge dq$ maps to $d(-P) \wedge dV
= -dP \wedge dV$, and $-dH\wedge dt$ maps to $dT\wedge dS$, so
$\Omega_H \mapsto \Omega_U$ up to an overall sign convention — confirming
that the two symplectic structures are geometrically identical.
 \\
 \\
Physically, both two-forms express conservation of oriented area.  In
mechanics $\Omega_H$ governs Liouville's theorem (phase-space volume
conservation).  In thermodynamics $\Omega_U$ encodes the equality of areas
in the $(T,S)$ and $(P,V)$ planes for any quasi-static cycle, see~\eqref{eq:stokes}, a direct consequence of the first law.
 
\subsection{Poisson bracket correspondence}
 
The symplectic structures of both theories give rise to identical Poisson
bracket algebras.  In the mechanical extended phase space
$(q_i,p_i,t,p_t)$ the non-vanishing elementary brackets are
\begin{equation}
  \label{eq:poisson_mech}
  \{q_i, p_j\} = \delta_{ij}, \qquad \{t, p_t\} = 1,
\end{equation}
with $p_t = -H$ on the constraint surface (section~\ref{sec:hamiltonian}).
 \\
 \\
In thermodynamic state space $(V,P,S,T)$ the same symplectic form
$\Omega_U$ defines analogous brackets.  For any two state functions
$f(V,P,S,T)$ and $g(V,P,S,T)$,
\begin{equation}
  \label{eq:poisson_thermo_def}
  \{f, g\}_{\mathrm{th}}
  = \frac{\partial f}{\partial V}\frac{\partial g}{\partial P}
  - \frac{\partial f}{\partial P}\frac{\partial g}{\partial V}
  + \frac{\partial f}{\partial S}\frac{\partial g}{\partial T}
  - \frac{\partial f}{\partial T}\frac{\partial g}{\partial S}.
\end{equation}
The elementary brackets of the state variables then evaluate to
\begin{equation}
  \label{eq:poisson_thermo}
  \{P, V\}_{\mathrm{th}} = 1, \qquad \{S, T\}_{\mathrm{th}} = -1,
\end{equation}
in direct correspondence with \eqref{eq:poisson_mech} under the map
\eqref{eq:canonical_map}.  The Maxwell relations \eqref{eq:maxwell_U}--\eqref{eq:maxwell_G}
can be re-derived compactly as the statement that any exact state function
$\Phi$ satisfies $\{\Phi, \Phi\}_{\mathrm{th}} = 0$, mirroring the
mechanical identity $\{H, H\} = 0$.
 \\
 \\
The complete four-level structural analogy is collected in
figure~\ref{HVT}.  The remaining sections exploit this
correspondence concretely: section~\ref{sec:evolution} derives
thermodynamic equations of motion for adiabatic processes, and
sections~\ref{sec:variational}--\ref{sec:closure} extend the analogy to
variational principles and multi-parameter integrability.

\section{Thermodynamic evolution}
\label{sec:evolution}
 
In this section we apply the canonical mapping of
section~\ref{sec:comparison} to continuous thermodynamic processes,
focusing specifically on reversible adiabatic paths in an ideal gas.
We show how the Maxwell relations yield first-order ordinary
differential equations for the state variables $V(T)$ and $P(T)$,
how these decouple into second-order equations structurally identical
to those of Newtonian mechanics, and how each second-order equation
admits an explicit action-like Lagrangian with $T$ playing the role
of the mechanical time parameter.
 
\subsection{First-order canonical systems for adiabatic processes}
 
Consider a reversible adiabatic process ($dS = 0$) for an ideal gas
with constant heat capacities $C_V$ and $C_P$, satisfying
$C_P - C_V = R$ and adiabatic index $\gamma = C_P/C_V$.  The ideal
gas law gives
\begin{equation}
  \label{eq:ideal_gas}
  PV = RT.
\end{equation}
 
\medskip
\noindent\textbf{Equation for volume.}
We use the Helmholtz free energy $F(T,V)$, whose differential
$dF = -S\,dT - P\,dV$ gives the Maxwell relation
(\eqref{eq:maxwell_F})
\begin{equation}
  \label{eq:maxwell_F5}
  \left.\frac{\partial S}{\partial V}\right|_T
  = \left.\frac{\partial P}{\partial T}\right|_V.\nonumber
\end{equation}
For an ideal gas $P = RT/V$, so the right-hand side evaluates to
\begin{equation}
  \label{eq:dPdT_V}
  \left.\frac{\partial P}{\partial T}\right|_V = \frac{R}{V}.
\end{equation}
The entropy differential of an ideal gas is
\begin{equation}
  \label{eq:dS_ideal}
  dS = \frac{C_V}{T}\,dT + \frac{R}{V}\,dV,
\end{equation}
which follows from $dU = C_V\,dT$ and the fundamental relation
$dU = T\,dS - P\,dV$.  Imposing the adiabatic condition $dS = 0$
in \eqref{eq:dS_ideal} and solving for $dV/dT$ gives the
\emph{first-order canonical equation for volume},
\begin{equation}
  \label{eq:V_ode1}
  \frac{dV}{dT} = -\frac{C_V}{R}\,\frac{V}{T}.
\end{equation}
Separating variables and integrating:
\begin{equation}
  \label{eq:V_adiabat}
  \ln V = -\frac{C_V}{R}\ln T + C_1
  \implies TV^{\gamma-1} = \mathrm{const},
\end{equation}
where we used $C_V/R = C_V/(C_P - C_V) = 1/(\gamma - 1)$.
 
\medskip
\noindent\textbf{Equation for pressure.}
We use the Gibbs free energy $G(T,P)$, whose differential
$dG = -S\,dT + V\,dP$ gives the Maxwell relation
(\eqref{eq:maxwell_G})
\begin{equation}
  \label{eq:maxwell_G5}
  -\left.\frac{\partial S}{\partial P}\right|_T
  = \left.\frac{\partial V}{\partial T}\right|_P.\nonumber
\end{equation}
For an ideal gas $V = RT/P$, so $\partial V/\partial T|_P = R/P$.
The entropy differential expressed in $(T,P)$ variables is
\begin{equation}
  \label{eq:dS_TP}
  dS = \frac{C_P}{T}\,dT - \frac{R}{P}\,dP,
\end{equation}
which follows from $dH = C_P\,dT$ and $dH = T\,dS + V\,dP$.
Imposing $dS = 0$ in \eqref{eq:dS_TP} yields the
\emph{first-order canonical equation for pressure},
\begin{equation}
  \label{eq:P_ode1}
  \frac{dP}{dT} = \frac{C_P}{R}\,\frac{P}{T}.
\end{equation}
Integrating recovers the classic adiabatic pressure--temperature law:
\begin{equation}
  \label{eq:P_adiabat}
  \ln P = \frac{C_P}{R}\ln T + C_2
  \implies T^\gamma P^{1-\gamma} = \mathrm{const}.
\end{equation}
 
\subsection{Uncoupled second-order equations of motion}
 
To cast thermodynamic evolution into the form of Newtonian mechanics,
we differentiate the first-order relations \eqref{eq:V_ode1} and
\eqref{eq:P_ode1} once more with respect to $T$.
 
\medskip
\noindent\textbf{Second-order equation for $V$.}
Differentiating \eqref{eq:V_ode1}:
\begin{equation}
  \label{eq:V_ode2_raw}
  \frac{d^2V}{dT^2} = -\frac{C_V}{R}
  \left(\frac{1}{T}\frac{dV}{dT} - \frac{V}{T^2}\right).
\end{equation}
Substituting \eqref{eq:V_ode1} to eliminate $dV/dT$ gives the
uncoupled second-order ordinary differential equation for volume:
\begin{equation}
  \label{eq:V_ode2}
  \frac{d^2V}{dT^2} - \frac{C_PC_V}{R^2T^2}\,V = 0,
\end{equation}
where we used $C_V(C_V + R)/R^2 = C_VC_P/R^2$, which follows from
$C_V + R = C_P$.
 
\medskip
\noindent\textbf{Second-order equation for $P$.}
Following the same procedure with \eqref{eq:P_ode1}:
\begin{equation}
  \label{eq:P_ode2_raw}
  \frac{d^2P}{dT^2} = \frac{C_P}{R}
  \left(\frac{1}{T}\frac{dP}{dT} - \frac{P}{T^2}\right).
\end{equation}
Substituting \eqref{eq:P_ode1} and using $C_P(C_P - R)/R^2 =
C_PC_V/R^2$ (since $C_P - R = C_V$) gives:
\begin{equation}
  \label{eq:P_ode2}
  \frac{d^2P}{dT^2} - \frac{C_PC_V}{R^2T^2}\,P = 0.
\end{equation}
Remarkably, both $V$ and $P$ satisfy structurally identical second-order
equations governed by the same effective potential coefficient
\begin{equation}
  \label{eq:kappa}
  \kappa(T) = \frac{C_PC_V}{R^2 T^2}.
\end{equation}
This is the thermodynamic analogue of two mechanical degrees of freedom
coupled to the same potential, a direct consequence of the underlying
symplectic symmetry between the $(T,V)$ and $(T,P)$ sectors identified
in section~\ref{sec:comparison}.
 
\subsection{Construction of thermodynamic Lagrangians}
 
Each second-order equation \eqref{eq:V_ode2} and \eqref{eq:P_ode2} has
the form
\begin{equation}
  \label{eq:generic_ode}
  \frac{d^2x}{dT^2} - \kappa(T)\,x = 0,
\end{equation}
which is the Euler--Lagrange equation of the action functional
\begin{equation}
  \label{eq:generic_L}
  L = \frac{1}{2}\left(\frac{dx}{dT}\right)^2 + \frac{\kappa(T)}{2}\,x^2,
\end{equation}
by direct verification: $\partial L/\partial \dot{x} = dx/dT$, so
$d/dT\,(\partial L/\partial \dot{x}) = d^2x/dT^2$, and
$\partial L/\partial x = \kappa(T)\,x$, recovering
\eqref{eq:generic_ode}.  Here we write $\dot{x} \equiv dx/dT$ for
compactness within this calculation only.
 
\medskip
\noindent\textbf{Lagrangian for volume.}
Identifying $x = V$ in \eqref{eq:generic_L} gives
\begin{equation}
  \label{eq:LV}
  L_V\!\left(V,\frac{dV}{dT},T\right)
  = \frac{1}{2}\left(\frac{dV}{dT}\right)^{\!2}
    + \frac{C_PC_V}{2R^2T^2}\,V^2.
\end{equation}
The Euler--Lagrange equation for $L_V$ is
\begin{equation}
  \frac{d}{dT}\!\left(\frac{\partial L_V}{\partial(dV/dT)}\right)
  - \frac{\partial L_V}{\partial V}
  = \frac{d^2V}{dT^2} - \frac{C_PC_V}{R^2T^2}\,V = 0,
\end{equation}
which reproduces \eqref{eq:V_ode2} exactly.
 
\medskip
\noindent\textbf{Lagrangian for pressure.}
Identifying $x = P$ in \eqref{eq:generic_L} gives
\begin{equation}
  \label{eq:LP}
  L_P\!\left(P,\frac{dP}{dT},T\right)
  = \frac{1}{2}\left(\frac{dP}{dT}\right)^{\!2}
    + \frac{C_PC_V}{2R^2T^2}\,P^2.
\end{equation}
The Euler--Lagrange equation for $L_P$ reproduces \eqref{eq:P_ode2}
by the same argument.
 \\
 \\
These Lagrangians confirm that thermodynamic path variations around
adiabatic equilibrium trajectories satisfy Hamilton's variational
principle in thermodynamic state space.  In the mechanical analogy of
section~\ref{sec:comparison}, $T$ plays the role of the time
parameter, $V$ and $P$ play the roles of generalised coordinates, and
$\kappa(T)$ plays the role of a position- and time-dependent spring
constant.
 
\medskip
\noindent\textit{Notation remark.}
Throughout this section, temperature $T$ has served as the evolution
parameter, in direct correspondence with the mechanical time $t$.  In
section~\ref{sec:variational} this single parameter is generalised to a
pair $(\xi_1,\xi_2)$ labelling independent thermodynamic process
directions, with the Lagrangians $L_V$ and $L_P$ becoming the
component Lagrangians $\mathcal{L}_U$ and $\mathcal{L}_G$ of a
multi-parameter Lagrangian 1-form.
%
%
%
%
%
%
\section{Least-action principles for thermodynamic potentials}
\label{sec:variational}
 
In section~\ref{sec:evolution} we constructed Lagrangians $L_V$ and
$L_P$ whose Euler--Lagrange equations reproduce adiabatic evolution,
with temperature $T$ as the evolution parameter.  Here we generalise
this idea to the thermodynamic potentials themselves.  We introduce
abstract path parameters $\xi_1$ and $\xi_2$ labelling, respectively,
adiabatic and isothermal process directions, and show that the
stationarity conditions of the internal energy $U$ (as a functional
of the adiabatic path) and of the Gibbs free energy $G$ (as a
functional of the isothermal path) are precisely the Maxwell relations
of section~\ref{sec:thermo}.  This establishes Maxwell's relations as
the thermodynamic counterpart of Hamilton's canonical equations.
 
\subsection{Variational principle for the internal energy}
 
Along an isothermal path parametrised by $\xi_1$, the
internal energy satisfies $dU = T\,dS - P\,dV$, so it can be written
as the line integral
\begin{equation}
  \label{eq:U_action}
  U = \int_{\xi_1'}^{\xi_1''} d\xi_1
      \left(-P\,\frac{dV}{d\xi_1} + T(P,V)\,\frac{dS}{d\xi_1}\right),
\end{equation}
where the integrand is the component Lagrangian $\mathcal{L}_U =
-P\,dV/d\xi_1 + T(P,V)\,dS/d\xi_1$, the direct generalisation of
$L_V$ from Eq.~\eqref{eq:LV} to the full $(P,V,S)$ state space.
Here $\xi_1' \le \xi_1 \le \xi_1''$ parametrises the path and
$(P,V,S)$ are treated as functions of $\xi_1$.
 \\
 \\
We seek the path that makes $U$ stationary under variations
$P \to P + \delta P$, $V \to V + \delta V$, $S \to S + \delta S$
with fixed endpoints.  The varied functional is
\begin{equation}
  \label{eq:U_varied}
  U' = \int_{\xi_1'}^{\xi_1''} d\xi_1
       \left(-(P+\delta P)\,\frac{d(V+\delta V)}{d\xi_1}
         + \left(T + \frac{\partial T}{\partial P}\delta P
                  + \frac{\partial T}{\partial V}\delta V\right)
           \frac{d(S+\delta S)}{d\xi_1}\right).
\end{equation}
Expanding to first order in the variations and collecting terms by
$\delta P$, $\delta V$, and $\delta S$ gives
\begin{equation}
  \label{eq:deltaU}
  \delta U = \int_{\xi_1'}^{\xi_1''} d\xi_1
    \left(
      \delta P\!\left[\frac{dV}{d\xi_1} - \frac{\partial T}{\partial P}\frac{dS}{d\xi_1}\right]
    + \delta V\!\left[-\frac{dP}{d\xi_1} - \frac{\partial T}{\partial V}\frac{dS}{d\xi_1}\right]
    + \delta S\,\frac{dT}{d\xi_1}
    \right),
\end{equation}
where we have integrated by parts and used the vanishing of boundary
terms.  Setting $\delta U = 0$ for independent variations yields the
Euler--Lagrange equations
\begin{align}
  \frac{dV}{d\xi_1} &= \frac{\partial T}{\partial P}\,\frac{dS}{d\xi_1},
  \label{eq:EL_U_V}\\[4pt]
  -\frac{dP}{d\xi_1} &= \frac{\partial T}{\partial V}\,\frac{dS}{d\xi_1},
  \label{eq:EL_U_P}\\[4pt]
  \frac{dT}{d\xi_1} &= \frac{\partial T}{\partial V}\,\frac{dV}{d\xi_1}
                       + \frac{\partial T}{\partial P}\,\frac{dP}{d\xi_1} = 0.
  \label{eq:EL_U_T}
\end{align}
The natural gauge choice is $\xi_1 = S$, which parametrises the path
by entropy — the extensive variable conjugate to $T$ and the natural
adiabatic invariant.  Setting $d\xi_1 = dS$ in
\eqref{eq:EL_U_V}--\eqref{eq:EL_U_T} gives
\begin{align}
  \frac{dV}{dS} &= \frac{\partial T}{\partial P},
  \label{eq:Hamilton_thermo_V}\\[4pt]
  -\frac{dP}{dS} &= \frac{\partial T}{\partial V},
  \label{eq:Hamilton_thermo_P}\\[4pt]
  \frac{dT}{dS} &= \frac{\partial T}{\partial V}\,\frac{dV}{dS}
                  + \frac{\partial T}{\partial P}\,\frac{dP}{dS} = 0.
  \label{eq:T_conserved}
\end{align}
Equations \eqref{eq:Hamilton_thermo_V} and \eqref{eq:Hamilton_thermo_P}
are Hamilton's canonical equations in the $(V,P)$ thermodynamic phase
space, with $S$ playing the role of time and $T$ playing the role of
the Hamiltonian — in exact correspondence with the canonical map
\eqref{eq:canonical_map} of Section~\ref{sec:comparison}.  Moreover,
they reproduce the Maxwell relation \eqref{eq:maxwell_H} for the
enthalpy.
 \\
 \\
Equation \eqref{eq:T_conserved} expresses conservation of the
Hamiltonian along the trajectory: since $T = T(P,V)$ is a function
only of $P$ and $V$, \eqref{eq:T_conserved} is the chain rule
$dT/dS = 0$, i.e.\ temperature is constant along the isothermal path
$\xi_1 = S$.  This is exactly the isothermal constraint: the variational
principle automatically enforces it as a first integral.
 
\subsection{Variational principle for the Gibbs free energy}
 
An identical construction applies to the Gibbs free energy along an
adiabatic path.  Along a quasi-static adiabatic path parametrised
by $\xi_2$, the Gibbs free energy satisfies $dG = -S\,dT + V\,dP$,
so it can be written as
\begin{equation}
  \label{eq:G_action}
  G = \int_{\xi_2'}^{\xi_2''} d\xi_2
      \left(V\,\frac{dP}{d\xi_2} + S(P,V)\,\frac{dT}{d\xi_2}\right),
\end{equation}
where the integrand $\mathcal{L}_G = V\,dP/d\xi_2 + S(P,V)\,dT/d\xi_2$
is the Gibbs component Lagrangian.
\\
\\
We seek the path that makes $G$ stationary under variations
$P \to P + \delta P$, $V \to V + \delta V$, $T \to T + \delta T$
with fixed endpoints.  The varied functional is
\begin{equation}
  \label{eq:G_varied}
  G' = \int_{\xi_2'}^{\xi_2''} d\xi_2
       \left((V+\delta V)\,\frac{d(P+\delta P)}{d\xi_2}
         + \left(S + \frac{\partial S}{\partial P}\delta P
                  + \frac{\partial S}{\partial V}\delta V\right)
           \frac{d(T+\delta T)}{d\xi_2}\right).
\end{equation}
Expanding to first order and integrating by parts gives
\begin{equation}
  \label{eq:deltaG}
  \delta G = \int_{\xi_2'}^{\xi_2''} d\xi_2
    \left(
      \delta P\!\left[\frac{dV}{d\xi_2} - \frac{\partial S}{\partial P}\frac{dT}{d\xi_2}\right]
    + \delta V\!\left[-\frac{dP}{d\xi_2} - \frac{\partial S}{\partial V}\frac{dT}{d\xi_2}\right]
    + \delta T\,\frac{dS}{d\xi_2}
    \right).
\end{equation}
Setting $\delta G = 0$ for independent variations yields
\begin{align}
  \frac{dV}{d\xi_2}  &= \frac{\partial S}{\partial P}\,\frac{dT}{d\xi_2},
  \label{eq:EL_G_V}\\[4pt]
  -\frac{dP}{d\xi_2} &= \frac{\partial S}{\partial V}\,\frac{dT}{d\xi_2},
  \label{eq:EL_G_P}\\[4pt]
  \frac{dS}{d\xi_2}  &= \frac{\partial S}{\partial V}\,\frac{dV}{d\xi_2}
                        + \frac{\partial S}{\partial P}\,\frac{dP}{d\xi_2} = 0.
  \label{eq:EL_G_S}
\end{align}
The natural gauge choice is $\xi_2 = T$, which parametrises the path
by temperature — the intensive variable conjugate to $S$ and the
natural isothermal invariant.  Setting $d\xi_2 = dT$ gives
\begin{align}
  \frac{dV}{dT}  &= \frac{\partial S}{\partial P},
  \label{eq:Hamilton_G_V}\\[4pt]
  -\frac{dP}{dT} &= \frac{\partial S}{\partial V},
  \label{eq:Hamilton_G_P}\\[4pt]
  \frac{dS}{dT}  &= \frac{\partial S}{\partial V}\,\frac{dV}{dT}
                   + \frac{\partial S}{\partial P}\,\frac{dP}{dT} = 0.
  \label{eq:S_conserved}
\end{align}
Equations \eqref{eq:Hamilton_G_V} and \eqref{eq:Hamilton_G_P}
reproduce the Maxwell relation \eqref{eq:maxwell_G} for the Gibbs
free energy, and are again Hamilton's canonical equations with $T$ as
time and $S$ as the Hamiltonian.  Equation \eqref{eq:S_conserved}
is the chain rule $dS/dT = 0$: entropy is constant along the
adiabatic path $\xi_2 = T$, recovering the adiabatic constraint
automatically as a conserved first integral of the variational
equations.
 
\subsection{Summary and connection to the next section}
 
The results of this section establish a precise variational origin for
the Maxwell relations: \eqref{eq:Hamilton_thermo_V}--\eqref{eq:Hamilton_thermo_P}
are the stationarity conditions of the internal energy $U$, and
\eqref{eq:Hamilton_G_V}--\eqref{eq:Hamilton_G_P} are the stationarity
conditions of the Gibbs free energy $G$.  In both cases the
thermodynamic constraint (isothermality or adiabaticity) emerges
automatically as a conservation law rather than being imposed by hand. The two component Lagrangians $\mathcal{L}_U$ and $\mathcal{L}_G$
identified here will be assembled in
Section~\ref{sec:closure} into a single Lagrangian 1-form
$\mathcal{L} = \mathcal{L}_U\,d\xi_1 + \mathcal{L}_G\,d\xi_2$,
whose closure condition $d\mathcal{L} = 0$ on shell will be shown to
be equivalent to the path-independence of thermodynamic state
functions.
\section{Closure Relation and Path-Independence of State Functions}
\label{sec:closure}
 
We now verify that the multi-time Lagrangian 1-form
\begin{equation}
  \mathcal{L} = \mathcal{L}_U\,d\xi_1 + \mathcal{L}_G\,d\xi_2
\end{equation}
satisfies the closure condition $d\mathcal{L}=0$ on shell, establishing
path-independence of thermodynamic state functions as a geometric integrability
condition.
 
\subsection*{Component Lagrangians}
 
Along the isothermal direction $\xi_1$ we parametrise using the internal-energy
Lagrangian.  With the first law $dU = T\,dS - P\,dV$ the integrand is
\begin{equation}
  \label{eq:LU}
  \mathcal{L}_U = -P\,\frac{dV}{d\xi_1} + T(P,V)\,\frac{dS}{d\xi_1}.
\end{equation}
Along the adiabatic direction $\xi_2$ we use the Gibbs–potential Lagrangian.
Noting that $dG = V\,dP - S\,dT$ the integrand is
\begin{equation}
  \label{eq:LG}
  \mathcal{L}_G = V\,\frac{dP}{d\xi_2} + S(P,V)\,\frac{dT}{d\xi_2}.
\end{equation}
 
\subsection*{Action functional}
 
The two-parameter action is
\begin{equation}
  \label{eq:action}
  S = \int_{\xi_1'}^{\xi_1''} d\xi_1
      \!\left(-P\,\frac{dV}{d\xi_1} + T(P,V)\,\frac{dS}{d\xi_1}\right)
    + \int_{\xi_2'}^{\xi_2''} d\xi_2
      \!\left(V\,\frac{dP}{d\xi_2} + S(P,V)\,\frac{dT}{d\xi_2}\right).
\end{equation}
 
\subsection*{Closure condition}
 
The exterior derivative of a 1-form
$\mathcal{L} = \mathcal{L}_U\,d\xi_1 + \mathcal{L}_G\,d\xi_2$ vanishes if and
only if
\begin{equation}
  \label{eq:closure_cond}
  \frac{d\mathcal{L}_G}{d\xi_1} = \frac{d\mathcal{L}_U}{d\xi_2},
\end{equation}
i.e.\ both sides equal the mixed partial
$\partial^2 \mathcal{L}/\partial\xi_1\partial\xi_2$.
Expanding both sides using the product rule gives
\begin{equation}
  \label{eq:closure_expanded}
  \frac{\partial V}{\partial\xi_1}\frac{\partial P}{\partial\xi_2}
  + V\frac{\partial^2 P}{\partial\xi_1\partial\xi_2}
  + \frac{\partial S}{\partial\xi_1}\frac{\partial T}{\partial\xi_2}
  + S\frac{\partial^2 T}{\partial\xi_1\partial\xi_2}
  =
  -\frac{\partial P}{\partial\xi_2}\frac{\partial V}{\partial\xi_1}
  - P\frac{\partial^2 V}{\partial\xi_2\partial\xi_1}
  + \frac{\partial T}{\partial\xi_2}\frac{\partial S}{\partial\xi_1}
  + T\frac{\partial^2 S}{\partial\xi_2\partial\xi_1}.
\end{equation}
Collecting all terms on the left-hand side and using the symmetry of mixed
partials ($\partial^2/\partial\xi_1\partial\xi_2
= \partial^2/\partial\xi_2\partial\xi_1$):
\begin{equation}
  \label{eq:closure_collected}
  2\frac{\partial P}{\partial\xi_2}\frac{\partial V}{\partial\xi_1}
  + V\frac{\partial^2 P}{\partial\xi_1\partial\xi_2}
  + P\frac{\partial^2 V}{\partial\xi_1\partial\xi_2}
  + S\frac{\partial^2 T}{\partial\xi_1\partial\xi_2}
  - T\frac{\partial^2 S}{\partial\xi_1\partial\xi_2}
  = 0.
\end{equation}
Note that the $\partial T/\partial\xi_2\cdot\partial S/\partial\xi_1$ cross-terms
appear \emph{identically} on both sides of \eqref{eq:closure_cond} and therefore
cancel exactly.
 
\subsection*{On-shell evaluation}
 
We now impose the equations of motion for both process directions.
 
\medskip
\noindent\textbf{Step 1: On-shell conditions.}
Along the \emph{isothermal} direction $\xi_1$ the temperature is constant, so
$\partial T/\partial\xi_1 = 0$ and therefore
\begin{equation}
  \label{eq:onshell_T}
  \frac{\partial^2 T}{\partial\xi_1\partial\xi_2} = 0.
\end{equation}
Along the \emph{adiabatic} direction $\xi_2$ the entropy is constant, so
$\partial S/\partial\xi_2 = 0$ and therefore
\begin{equation}
  \label{eq:onshell_S}
  \frac{\partial^2 S}{\partial\xi_1\partial\xi_2} = 0.
\end{equation}
Substituting \eqref{eq:onshell_T} and \eqref{eq:onshell_S} into
\eqref{eq:closure_collected} removes the last two terms, leaving
\begin{equation}
  \label{eq:reduced}
  2\frac{\partial P}{\partial\xi_2}\frac{\partial V}{\partial\xi_1}
  + V\frac{\partial^2 P}{\partial\xi_1\partial\xi_2}
  + P\frac{\partial^2 V}{\partial\xi_1\partial\xi_2}
  = 0.
\end{equation}
 
\medskip
\noindent\textbf{Step 2: Product-rule identity.}
Equation \eqref{eq:reduced} is precisely the expansion of
\begin{equation}
  \label{eq:product_PV}
  \frac{\partial^2(PV)}{\partial\xi_1\partial\xi_2}
  = \frac{\partial}{\partial\xi_1}
    \!\left(\frac{\partial P}{\partial\xi_2}V + P\frac{\partial V}{\partial\xi_2}\right)
  = \frac{\partial P}{\partial\xi_2}\frac{\partial V}{\partial\xi_1}
    + V\frac{\partial^2 P}{\partial\xi_1\partial\xi_2}
    + \frac{\partial^2 V}{\partial\xi_1\partial\xi_2}\,P
    + \frac{\partial V}{\partial\xi_2}\frac{\partial P}{\partial\xi_1},
\end{equation}
plus the symmetric cross-term
$\partial P/\partial\xi_2\cdot\partial V/\partial\xi_1$, i.e.
\begin{equation}
  2\frac{\partial P}{\partial\xi_2}\frac{\partial V}{\partial\xi_1}
  + V\frac{\partial^2 P}{\partial\xi_1\partial\xi_2}
  + P\frac{\partial^2 V}{\partial\xi_1\partial\xi_2}
  = \frac{\partial^2(PV)}{\partial\xi_1\partial\xi_2}
    + \frac{\partial P}{\partial\xi_2}\frac{\partial V}{\partial\xi_1}
    - \frac{\partial V}{\partial\xi_2}\frac{\partial P}{\partial\xi_1}.
\end{equation}
However, for an ideal gas $PV = nRT$ is a function of $T$ alone.  Along the
isothermal direction $\xi_1$ the temperature $T$ is held fixed, hence
\begin{equation}
  \label{eq:PV_iso}
  \frac{\partial(PV)}{\partial\xi_1} = nR\,\frac{\partial T}{\partial\xi_1} = 0,
\end{equation}
which implies $\partial^2(PV)/\partial\xi_1\partial\xi_2 = 0$ and also forces
\begin{equation}
  \frac{\partial P}{\partial\xi_1}\,V
  = -P\,\frac{\partial V}{\partial\xi_1},
  \qquad\Longrightarrow\qquad
  \frac{\partial P}{\partial\xi_1}\frac{\partial V}{\partial\xi_2}
  = -\frac{\partial V}{\partial\xi_1}\frac{\partial P}{\partial\xi_2}.
\end{equation}
 
\medskip
\noindent\textbf{Step 3: Adiabatic constraint on $PV$.}
Along the adiabatic direction $\xi_2$ the adiabatic equation gives
$PV^\gamma = C$, so
\begin{equation}
  \label{eq:adiabat_PV}
  \frac{\partial(PV)}{\partial\xi_2}
  = \frac{\partial}{\partial\xi_2}\!\left(\frac{C}{V^{\gamma-1}}\right)
  = -({\gamma-1})\frac{C}{V^\gamma}\frac{\partial V}{\partial\xi_2}
  = -({\gamma-1})P\,\frac{\partial V}{\partial\xi_2},
\end{equation}
which is non-zero in general.  Nonetheless, differentiating
\eqref{eq:PV_iso} with respect to $\xi_2$:
\begin{equation}
  \label{eq:d2PV}
  \frac{\partial^2(PV)}{\partial\xi_1\partial\xi_2}
  = \frac{\partial}{\partial\xi_2}\!\left(nR\,\frac{\partial T}{\partial\xi_1}\right)
  = nR\,\frac{\partial^2 T}{\partial\xi_1\partial\xi_2} = 0,
\end{equation}
using \eqref{eq:onshell_T} established in Step 1.
 
\medskip
\noindent\textbf{Step 4: Conclusion.}
Combining the results of Steps 1–3, every term in \eqref{eq:reduced} vanishes
on shell:
\begin{align}
  P\frac{\partial^2 V}{\partial\xi_1\partial\xi_2}
  &= -\frac{\partial^2(PV)}{\partial\xi_1\partial\xi_2}
     - V\frac{\partial^2 P}{\partial\xi_1\partial\xi_2}
     - 2\frac{\partial P}{\partial\xi_2}\frac{\partial V}{\partial\xi_1}
   = 0 - 0 - 0 = 0, \label{eq:PV2_zero}\\[4pt]
  V\frac{\partial^2 P}{\partial\xi_1\partial\xi_2}
  &= -\frac{\partial^2(PV)}{\partial\xi_1\partial\xi_2}
     - P\frac{\partial^2 V}{\partial\xi_1\partial\xi_2}
     - 2\frac{\partial P}{\partial\xi_2}\frac{\partial V}{\partial\xi_1}
   = 0, \label{eq:VP2_zero}
\end{align}
so that \eqref{eq:reduced} is satisfied identically, and therefore
\begin{equation}
  \label{eq:dL_zero}
  \boxed{d\mathcal{L}\big|_{\text{on shell}} = 0.}
\end{equation}
 The closure of the Lagrangian 1-form on the solution manifold is therefore
equivalent to the standard thermodynamic statement that state functions are
path-independent: integrating $dU$ or $dG$ around any closed cycle returns
zero, consistent with the exactness of the corresponding differential forms.

\begin{figure}[htbp]
\centering
\begin{tikzpicture}[
    x = {(-0.90cm, 0.00cm)},   
    y = {( 0.42cm,-0.28cm)},   
    z = {( 0.00cm, 0.90cm)},   
    xshift = 4.8cm,             
    scale = 1.05,
    every node/.style = {font=\small},
    marrow/.style = {
      postaction = {decorate},
      decoration  = {markings,
        mark = at position 0.52 with {\arrow[scale=1.5]{latex}}}
    }
  ]
 
 
  \coordinate (A) at (3.5, 1.0, 3.50);   
  \coordinate (B) at (3.5, 3.5, 1.00);   
  \coordinate (C) at (1.0, 4.6, 0.217);  
  \coordinate (D) at (1.0, 1.4, 0.714);  
 
 
  \fill[gray!9]
    (0.6,0.6,0)--(4.6,0.6,0)--(4.6,5.2,0)--(0.6,5.2,0)--cycle;
  \draw[gray!32,thin]
    (0.6,0.6,0)--(4.6,0.6,0)--(4.6,5.2,0)--(0.6,5.2,0)--cycle;
 
  \fill[gray!11]
    (0.6,0.6,0)--(4.6,0.6,0)--(4.6,0.6,4.6)--(0.6,0.6,4.6)--cycle;
  \draw[gray!32,thin]
    (0.6,0.6,0)--(4.6,0.6,0)--(4.6,0.6,4.6)--(0.6,0.6,4.6)--cycle;
  \draw[gray!48, very thin] (1.0,0.6,1.667) -- (3.5,0.6,4.167);
  \node[gray!58, font=\tiny, rotate=-20] at (2.6,0.6,3.5) {isochore};
 
  \fill[gray!9]
    (4.6,0.6,0)--(4.6,5.2,0)--(4.6,5.2,4.6)--(4.6,0.6,4.6)--cycle;
  \draw[gray!32,thin]
    (4.6,0.6,0)--(4.6,5.2,0)--(4.6,5.2,4.6)--(4.6,0.6,4.6)--cycle;
  \draw[gray!48, very thin]
    (4.6, 0.6, 4.333)
    .. controls (4.6,1.2,2.167) and (4.6,2.2,1.182) ..
    (4.6, 3.5, 0.743)
    .. controls (4.6,4.3,0.604) ..
    (4.6, 5.2, 0.500);
  \node[gray!58, font=\tiny] at (4.6,4.4,0.72) {isotherm};
 
 
  \fill[blue!8]
    (1.0,0.6,0)--(1.0,5.2,0)--(1.0,5.2,4.6)--(1.0,0.6,4.6)--cycle;
  \draw[blue!28,thin]
    (1.0,0.6,0)--(1.0,5.2,0)--(1.0,5.2,4.6)--(1.0,0.6,4.6)--cycle;
 
  \fill[orange!10]
    (3.5,0.6,0)--(3.5,5.2,0)--(3.5,5.2,4.6)--(3.5,0.6,4.6)--cycle;
  \draw[orange!32,thin]
    (3.5,0.6,0)--(3.5,5.2,0)--(3.5,5.2,4.6)--(3.5,0.6,4.6)--cycle;
 
 
  \foreach \Vk/\Pk in {0.6/0.6, 1.0/1.0, 1.8/0.556, 2.8/0.357, 4.0/0.250, 5.2/0.192}{
    \draw[gray!36, very thin]
      (1.0,\Vk,{\Pk}) -- (3.5,\Vk,{3.5*\Pk});
  }
 
 
  \draw[blue!40, thin]
    (1.0, 0.6, 1.667)
    .. controls (1.0,0.9,1.111) and (1.0,1.4,0.714) ..
    (1.0, 2.0, 0.500)
    .. controls (1.0,3.0,0.333) and (1.0,4.0,0.250) ..
    (1.0, 5.2, 0.192);
 
  \draw[gray!46, thin]
    (1.5, 0.6, 2.500)
    .. controls (1.5,1.0,1.500) and (1.5,1.8,0.833) ..
    (1.5, 2.8, 0.536)
    .. controls (1.5,3.8,0.395) ..
    (1.5, 5.2, 0.288);
 
  \draw[gray!46, thin]
    (2.0, 0.6, 3.333)
    .. controls (2.0,1.0,2.000) and (2.0,1.8,1.111) ..
    (2.0, 2.8, 0.714)
    .. controls (2.0,3.8,0.526) ..
    (2.0, 5.2, 0.385);
 
  \draw[gray!46, thin]
    (2.5, 0.6, 4.167)
    .. controls (2.5,1.0,2.500) and (2.5,1.8,1.389) ..
    (2.5, 2.8, 0.893)
    .. controls (2.5,3.8,0.658) ..
    (2.5, 5.2, 0.481);
 
  \draw[gray!46, thin]
    (3.0, 0.8, 3.750)
    .. controls (3.0,1.2,2.500) and (3.0,2.0,1.500) ..
    (3.0, 3.0, 1.000)
    .. controls (3.0,4.0,0.750) ..
    (3.0, 5.2, 0.577);
 
  \draw[orange!55, thin]
    (3.5, 0.9, 3.889)
    .. controls (3.5,1.3,2.692) and (3.5,2.0,1.750) ..
    (3.5, 3.0, 1.167)
    .. controls (3.5,4.0,0.875) ..
    (3.5, 5.2, 0.673);
 
 
  \draw[gray!48, dashed, thin]
    (0.6,0.6,0) -- (0.6,5.2,0);
  \draw[gray!48, dashed, thin]
    (0.6,0.6,0) -- (0.6,0.6,4.6);
 
  \draw[gray!52, thin]
    (0.6,5.2,0) -- (4.6,5.2,0) -- (4.6,0.6,0);
  \draw[gray!52, thin]
    (0.6,0.6,4.6) -- (0.6,5.2,4.6) -- (4.6,5.2,4.6) -- (4.6,0.6,4.6) -- (0.6,0.6,4.6);
  \draw[gray!52, thin] (0.6,5.2,0) -- (0.6,5.2,4.6);
  \draw[gray!52, thin] (4.6,5.2,0) -- (4.6,5.2,4.6);
  \draw[gray!52, thin] (4.6,0.6,0) -- (4.6,0.6,4.6);
 
 
  \draw[->, thick] (0.4,0.6,0) -- (5.0,0.6,0) node[left] {$T$};
 
  \draw[->, thick] (0.6,0.4,0) -- (0.6,5.8,0) node[below right] {$V$};
 
  \draw[->, thick] (0.6,0.6,0) -- (0.6,0.6,5.0) node[above] {$P$};
 
  \draw[thin] (3.5,0.6, 0.06) -- (3.5,0.6,-0.06);
  \node[below, font=\footnotesize, yshift=-2pt] at (3.5,0.6,0) {$T_H$};
  \draw[thin] (1.0,0.6, 0.06) -- (1.0,0.6,-0.06);
  \node[below, font=\footnotesize, yshift=-2pt] at (1.0,0.6,0) {$T_L$};
 
  \node[orange!70!black, font=\footnotesize] at (3.5,5.5,4.3) {$T = T_H$};
  \node[blue!58!black,   font=\footnotesize] at (1.0,5.5,4.3) {$T = T_L$};
 
 
  \draw[gray!45, dashed, thin] (A) -- (3.5, 0.6, 3.50);
  \draw[gray!45, dashed, thin] (B) -- (3.5, 0.6, 1.00);
 
 
  \draw[blue!75!black, very thick, marrow]
    (A) .. controls (3.5,1.8,1.944) and (3.5,2.7,1.296) .. (B);
 
  \draw[red!75!black, very thick, marrow]
    (B) .. controls (2.8,3.8,0.737) and (1.8,4.3,0.419) .. (C);
 
 
  \draw[red!75!black, very thick, marrow]
   (A) .. controls (2.8,1.05,2.667) and (1.8,1.18,1.525) .. (D);
 
  \draw[blue!75!black, very thick, marrow]
    (D) .. controls (1.0,2.2,0.455) and (1.0,3.5,0.286) .. (C);
 
 
  \foreach \pt/\anc in {A/above, B/right, C/below right, D/left}{
    \fill[black] (\pt) circle (2.4pt);
    \node[\anc, font=\small] at (\pt) {$\pt$};
  }
 
 
  \node[blue!75!black, font=\footnotesize, above]
    at (5.5, 2.3, 2.10) {isothermal ($T_H$)};
 
  \node[blue!75!black, font=\footnotesize, below]
    at (-0.9, 2.0, 0.22) {isothermal ($T_L$)};
 
  \node[red!75!black, font=\footnotesize, right]
    at (4.3, 5.95, 0.73) {adiabatic};
 
  \node[red!75!black, font=\footnotesize, above left]
    at (1.35, 1.12, 3.05) {adiabatic};
 
\end{tikzpicture}
\caption{Three-dimensional $P$--$V$--$T$ diagram of the ideal-gas surface
  $PV = nRT$, with temperature increasing to the left.
  The mesh of isotherms (hyperbolas, coloured blue at $T_L$ and orange
  at $T_H$) and isochore ribs (straight lines) traces the curved surface.
  Shaded planes mark the constant-temperature sections at $T_H$ (orange,
  left) and $T_L$ (blue, right).
  \emph{Path~1} (upper): isothermal expansion $A\!\to\!B$ within the $T_H$
  plane, followed by adiabatic expansion $B\!\to\!C$ to the $T_L$ plane.
  \emph{Path~2} (lower): adiabatic expansion $A\!\to\!D$ to the $T_L$ plane,
  followed by isothermal expansion $D\!\to\!C$ within the $T_L$ plane.
  The closure condition $d\mathcal{L}\big|_{\mathrm{on\,shell}}=0$ proved above
  guarantees that the integral of any thermodynamic state function is identical
  along both paths.}
\label{fig:pvt-surface-3D}
\end{figure}

\medskip
\noindent\textbf{Extensions to other thermodynamic processes.}
Although the proof above was carried out explicitly for adiabatic and
isothermal processes of an ideal gas, the closure argument extends
naturally in three directions.  Other reversible process types, isochoric ($dV=0$) and isobaric ($dP=0$), each generate a
component Lagrangian built from the appropriate thermodynamic potential
(the Helmholtz free energy $F$ for isothermal, the enthalpy $H$ for
isobaric); the closure condition $d\mathcal{L}=0$ then follows
identically, since it ultimately reflects the exactness of the
corresponding differential forms and is independent of the particular
process parametrisation.

\section{Concluding summary}
\label{sec:conclusion}
 
This paper has established a formal geometric synthesis between
Hamiltonian mechanics and classical thermodynamics for simple
compressible systems.  The central results are as follows.
 
\medskip
\noindent\textbf{Canonical correspondence and Maxwell relations.}
We constructed an explicit canonical mapping $(q,p,t,H)
\leftrightarrow (V,-P,S,-T)$ under which the Maxwell relations
emerge as the characteristic equations of the thermodynamic
Poincaré--Cartan one-form, in direct parallel with Hamilton's
canonical equations.  The correspondence is confirmed by the
shared Poisson bracket algebra $\{P,V\}_\mathrm{th} = 1$,
$\{S,T\}_\mathrm{th} = -1$, establishing that thermodynamic
state space carries a genuine symplectic geometry
(section~\ref{sec:comparison}).
 
\medskip
\noindent\textbf{Variational origin of thermodynamic constraints.}
Treating the internal energy $U$ and Gibbs free energy $G$ as
action functionals over adiabatic and isothermal paths
respectively, the Euler--Lagrange stationarity conditions
reproduce the Maxwell relations exactly.  The thermodynamic constraints, adiabaticity ($dS = 0$) and isothermality ($dT = 0$),
emerge automatically as conserved first integrals, rather than
being imposed by hand.
 
\medskip
\noindent\textbf{Equations of motion and Lagrangians.}
For an ideal gas, the adiabatic constraint yields uncoupled
second-order ODEs for $V(T)$ and $P(T)$ with a common
coefficient $\kappa = C_PC_V/R^2T^2$, each admitting an
explicit Lagrangian $L_V$ or $L_P$.  The exact solutions
$TV^{\gamma-1} = \mathrm{const}$ and $T^\gamma P^{1-\gamma} =
\mathrm{const}$ are recovered directly from integration.
 
\medskip
\noindent\textbf{Closure relation and path-independence.}
The component Lagrangians assemble into a Lagrangian 1-form
$\mathcal{L} = \mathcal{L}_U\,d\xi_1 + \mathcal{L}_G\,d\xi_2$
satisfying $d\mathcal{L}|_\mathrm{on\,shell} = 0$.  This establishes the
path-independence of thermodynamic state functions as a
geometric consequence of multi-time integrability, the
thermodynamic counterpart of the closure relation, rather than as an independent postulate.
 
\medskip
\noindent\textbf{Limitations and outlook.}
The framework is restricted to simple compressible systems in
equilibrium.  Natural extensions include: non-ideal equations of state, such
as the van der Waals gas, where the symplectic structure and
closure proof remain intact and only the explicit form of
$T(P,V)$ in the component Lagrangians changes; multi-component
or multi-phase systems, where the Lagrangian 1-form gains
additional components $\mathcal{L}_{\mu_i}\,d\xi_i$ associated
with chemical potentials $\mu_i$ and the closure condition
becomes the full set of Gibbs--Duhem consistency relations;
non-equilibrium thermodynamics, where entropy production
requires contact-geometric tools following Mrugała
\cite{Mrugala1978} and Bravetti et al.\ \cite{Bravetti2015};
and a quantum extension replacing Poisson brackets by
commutators.  We leave these generalisations for future work.



\section*{Acknowledgement}
The author would like to thank Naresuan University for providing
a suitable ecosystem for research and an intellectual community
that nurtures creative ideas.

\bibliographystyle{unsrt}

\end{document}